\documentstyle[twocolumn,prb,aps,epsf]{revtex}

\begin{document}
\draft

\wideabs{

\title{Combined effect of Zeeman splitting and spin-orbit interaction on the Josephson current
in a S-2DEG-S structure}
\author{E. V. Bezuglyi$^{a,b}$, A. S. Rozhavsky$^{a,c}$, I. D. Vagner$^{c}$,
and P. Wyder$^{c}$}

\address{$^a$B.Verkin Institute for Low temperature Physics and
Engineering, 47 Lenin Avenue, 61164 Kharkov, Ukraine}
\address{$^b$Chalmers University of Technology, S-41296,
G\"oteborg, Sweden}
\address{$^c$Grenoble High Magnetic Field Laboratory, Max-Planck-Institut
f\"ur Festk\"orperforschung and Centre National de la Recherche
Scientifique, B.P.166, 38042 Grenoble Cedex 9, France}

\maketitle
\begin{abstract}
We analyze new spin effects in current-carrying state of
superconductor-2D electron gas-su\-perconductor (S-2DEG-S) device
with spin-polarized nuclei in 2DEG region. The hyperfine
interaction of 2D electrons with nuclear spins, described by the
effective magnetic field $\bbox{B}$, produces Zeeman splitting of
Andreev levels without orbital effects, that leads to the
interference pattern of supercurrent oscillations over $B$. The
spin-orbit effects in 2DEG cause strongly anisotropic dependence
of the Josephson current on the direction of $\bbox{B}$, which may
be used as a probe for the spin-orbit interaction intensity. Under
certain conditions, the system reveals the properties of
$\pi$-junction.
\end{abstract}

}

The spin-orbit (SO) and hyperfine (HF) interactions in GaAs
heterojunctions and similar 2D quantum Hall systems attract
permanent theoretical and experimental attention. The hyperfine
field of the nuclear spin subsystem acting upon the spins of
charge carriers may reach $10^4$ G.\cite{Wald} At low
temperatures, the nuclear spin relaxation time can be
macroscopically long,\cite{Dyak} so the nonequilibrium spin
population in heterojunctions, once created, is conserved during
hundreds of seconds.\cite{Berg} The Zeeman splitting combined with
a strong spin-orbit coupling in GaAs/AlGaAs 2DEG gives rise to a
novel class of coherent phenomena, e.g., the spontaneous
Aharonov-Bohm effect.\cite{Vagner}

\begin{figure}
\epsfxsize=8.5cm\epsffile{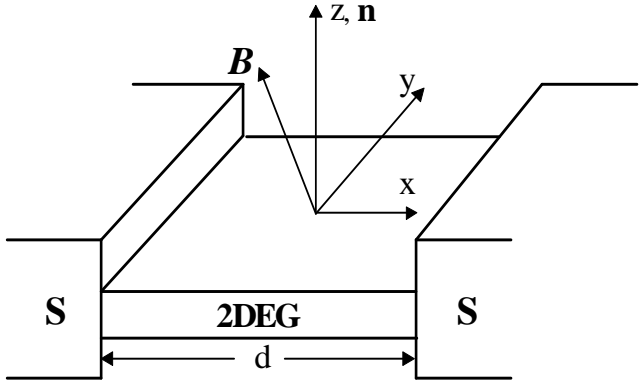} \vspace{3mm} \caption{A
model of the superconductor - 2DEG - superconductor device based
on GaAs/AlGaAs heterojunction.\cite{Taka} The normal $\bbox{n}$ is
directed towards the Al-doped layer.}
\end{figure}

In this paper we discuss mesoscopic spin-orbit effects in
Josephson current flowing across the S-2DEG-S structure (Fig.\ 1)
with polarized nuclei in 2DEG region. The transfer of the
Josephson current through the normal conducting layer is provided
by the Andreev reflection of electrons and holes at the
NS-interfaces, which convert normal electron excitations into
Cooper pairs in the superconducting banks. In a pure system with
length $d$ smaller than the electron scattering length, the
interference between coherent electron states and retro-reflected
hole states produces the set of spin-degenerate Andreev energy
levels $E_\lambda(\Phi)$, which depend on the quantum numbers
$\lambda$ and on the difference $\Phi$ of the order parameter
phases in superconducting electrodes.\cite{Kulik} In short
structures, $d \ll \xi_0$ ($\xi_0$ is the coherence length in the
superconductor), the Josephson current can be presented as the sum
of currents transferred by individual Andreev bound states (see,
e.g., Ref.\ \onlinecite{Shum} and references therein),
\begin{equation}
I(\Phi)=-{2e\over\hbar}\sum\nolimits_\lambda {\partial E_\lambda(\Phi)\over
\partial\Phi}\tanh{E_\lambda(\Phi)\over\ 2T},
\end{equation}
and must be sensitive to the HF and SO interaction which eliminate
spin degeneracy of the Andreev levels.

The contact HF interaction in a semiconductor is described by the
Hamiltonian\cite{Schlichter}
\begin{equation}
\hat{H}_{hf}=(8\pi/3)\mu_B\gamma_h\sum\nolimits_i\bbox{I}_i
\bbox{\sigma}\delta (\bbox{r}-\bbox{R}_i).
\end{equation}
Here $\mu_B$ is the Bohr magneton, $\gamma_h$ is the nuclear
magneton, and $\bbox{I}_i$, $\bbox{\sigma}$, $\bbox{R}_i$,
$\bbox{r}$ are nuclear and charge carrier spins and positions,
respectively. It follows from Eq.\ (2) that if the nuclear spins
are polarized, $\left\langle\sum\nolimits_i\bbox{I}_i
\right\rangle \neq 0$, the charge carrier spins feel the effective
HF field $\bbox{B}$ which may cause spin splitting in 2DEG of the
order of one tenth of the Fermi energy $E_F$.\cite{Wald,Berg} The
influence of the Zeeman splitting solely on a supercurrent was
studied first in Ref.\ \onlinecite{BB} for the SFS junction (F
denotes ferromagnetic metal). It was shown that the spin splitting
suppresses the critical current and produces its oscillations over
the intrinsic magnetic field localized within the F-layer.

The SO interaction of a charge carrier with the interface
potential in GaAs/AlGaAs heterojunctions is modelled by the
Bychkov-Rashba term,\cite{Bychkov,Pfeiffer}
\begin{equation}
\hat{H}_{\text{so}}=(\alpha/\hbar)\left[ \bbox{\sigma}
\bbox{p}\right] \bbox{n},  \label{Hso}
\end{equation}
where $\alpha =0.6\times 10^{-9}$ eV cm for
holes,\cite{Bychkov,Stormer} and $\alpha=0.25\times 10^{-9}$ eV cm
for electrons,\cite{Bychkov,Stein} $\bbox{\sigma}_i$, $\bbox{p}_i$
are the charge carrier spin and momentum, $\bbox{n}$ is the normal
to the interface directed towards the Al-doped layer.

The configuration proposed in Fig.\ 1 has the following
characteristic features:

i) the nuclear field $\bbox{B}$ is localized outside the
superconductors and does not influence the pairing mechanism;

ii) it affects only the electron spins and does not modify the
space motion of charge carriers, whereas usual magnetic field
causes strong orbital effects and transforms Andreev levels into
Landau bands;\cite{GB}

iii) since the SO interaction in Eq.\ (3) is spatially uniform
along the 2DEG plane, it does not suppress the supercurrent, in
contrast to the SO scattering at nonmagnetic impurities which acts
as a depairing factor.\cite{AG} The situation changes in presence
of the nuclear field.

To calculate the Josephson current in Eq.\ (1), it is enough to
find discrete eigenvalues of the BCS Hamiltonian supplied by the
HF and SO interaction terms,
\begin{equation}
{\cal H} =\!\!\int\!\! dV\!
\left[\psi^+\!\!\left(E(\hat{\bbox{p}})\! +\!
\bbox{\sigma}\hat{\bbox{\Lambda}}(\bbox{r})\right) \! \psi +
\Delta(\bbox{r})\psi_\uparrow^+\psi_\downarrow^+ \!+ \!h.c.\right]
\end{equation}
where $\bbox{\sigma}$ is the vector of Pauli matrices, the spinor
$\psi$ is composed from the annihilation operators
$\psi_\uparrow$, $\psi_\downarrow$ of the electron with a given
spin, and $E(\bbox{p}) = \bbox{p}^2/2m - E_F$ is the energy of a
normal electron excitation. For simplicity, we accept a step-wise
model for the order parameter $\Delta(\bbox{r})$,
\begin{equation}
\Delta(x) = \left\{\begin{array}{ccc}
\Delta\exp\left(i\Phi\,\mbox{sign}\,x/2\right), & |x|>d/2,\\ 0, &
|x|<d/2.
\end{array}\right.
\end{equation}

The vector $\hat{\bbox{\Lambda}}(\bbox{r})$ in Eq.\ (4) describes
the HF and SO interactions in accordance with Eqs.\ (2) and (3),
\begin{equation}
\hat{\bbox{\Lambda}}(x)\! =\! \left\{\begin{array}{ccc} 0, & |x|>d/2, \\
\bbox{h} + \hat{\bbox{w}}, & |x|<d/2,
\end{array}\right.\;
\bbox{h}\! = \!\mu_B\bbox{B}, \; \hat{\bbox{w}}\! \!=
{\alpha\over\hbar} [\hat{\bbox{p}} \bbox{n}].
\end{equation}

In order to consider uniformly the spin and electron-hole states
of quasiparticles, it is convenient to express the Hamiltonian in
Eq.\ (4) in terms of the 4-spinor field,
\begin{equation}
\varphi = \left( \begin{array}{ccc} \psi_\uparrow \\ \psi_\downarrow \\
\psi_\downarrow^+ \\ \psi_\uparrow^+
\end{array} \right), \qquad
\varphi^+ = \left(\psi_\uparrow^+, \; \psi_\downarrow^+, \; \psi_\downarrow,
\; \psi_\uparrow\right),
\end{equation}
\begin{equation}
{\cal H} = {1\over 2}\int dV
\varphi^+(\bbox{r})\hat{H}\varphi(\bbox{r}),
\end{equation}
\begin{equation}
\hat{H} = \tau_z E(\hat{\bbox{p}}) +\! \left\{
\begin{array}{lll}
\sigma_z[\Delta(x)\tau_+\! +\! \Delta^\ast(x)\tau_-],
& |x| > d/2,\\
\hat{\bbox{w}} \bbox{\sigma} +h_z\sigma_z +\tau_z
\bbox{h}_\parallel\bbox{\sigma}, & |x| < d/2.
\end{array}\right.
\end{equation}
Here $\bbox{h}_\parallel$ is the component of Zeeman field in 2DEG
plane; the Pauli matrices $\sigma$ act upon the spin states, while
$\tau$-matrices operate in electron-hole space, e.g.,
\begin{equation}
\sigma_z\tau_z = \left(\begin{array}{ccc}
\sigma_z & 0 \\ 0 & -\sigma_z
\end{array}\right).
\end{equation}

The problem of single-particle spectrum of the Hamiltonian
$\cal{H}$ in Eq.\ (8) is equivalent to the solution of the
Bogolyubov-de Gennes (BdG) equation for the 4-component wave
function $\Psi_\lambda(\bbox{r})$
\begin{equation}
\hat{H}\Psi_\lambda(\bbox{r}) = E_\lambda\Psi_\lambda(\bbox{r}).
\end{equation}

Assuming all characteristic energies $\Delta$, $h$, $w$ to be much
smaller than $E_F$, we use the quasiclassical representation of
$\Psi_\lambda(\bbox{r})$ as the product of rapidly oscillating
exponent over the slowly varying envelope $u(x)$:
\begin{equation}
\Psi_\lambda(\bbox{r}) = \exp(ispx + ip_y y) u_s(x),\;
p=\sqrt{p_F^2-p_y^2}
\end{equation}
where $s=\pm 1$ indicates two signs of $x$-component of the
electron momentum. The spinor function $u(x)$ obeys Eq.\ (11) with
a reduced Hamiltonian, in which the quasiclassical approximation
for the kinetic energy and SO operators is used,
\begin{equation}
E(\hat{\bbox{p}})\approx sv\hat{p}_x,\; \hat{\bbox{w}} \approx
{\alpha\over\hbar} [\bbox{pn}], \; \bbox{p} =(sp,p_y,0), \; v =
{p\over m}.
\end{equation}

The matching of the solutions of Eq.\ (11) at the NS interfaces,
which are assumed to be completely transparent, yields a
dispersion relation,
\begin{equation}
{E d\over\hbar v}\! =\! \pi n +\arccos{E\over\Delta}+ s{\Phi\over 2}
+ \sigma\gamma,
\;\; n = 0,\pm 1, \pm 2...,
\end{equation}
\begin{equation}
\gamma(\bbox{h},\bbox{w})\! =\!
\arcsin\!\!\left[\!\sum_{k=\pm1}\!\!{1\!+\!k\bbox{n}_+ \bbox{n}_-
\over 2} \sin^2\! {A_+\! +\! kA_-\over 2}\!\right]^{1/2}
\end{equation}
\begin{equation}
A_\pm = (d/\hbar v)|\bbox{h} \pm \bbox{w}|, \quad \bbox{n}_\pm =
(\bbox{h} \pm \bbox{w}) / |\bbox{h} \pm \bbox{w}|
\end{equation}
which has the standard structure of the equation for Andreev
levels\cite{foot1} in current-carrying SNS junction.\cite{Kulik}
An additional term $\sigma\gamma$, where $\sigma = \pm 1$
indicates the spin direction, describes Zeeman splitting of the
Andreev levels renormalized by the SO interaction. In terms of the
BdG wave mechanics, the spin effects change phase relations
between the wave functions of the incident and retro-reflected
quasiparticles. This produces oscillations of the Andreev levels
with the change of the interaction parameters $h$, $w$ which enter
the oscillating phase shift $\sigma\gamma$ in Eq.\ (14). As a
result, the Josephson current in Eq.\ (1) also reveals
oscillations with $h$, $w$ as the manifestation of complicated
interference between partial supercurrents carried by individual
Andreev bands. In this sense, the effect represents a spin-induced
analogue of the Fraunhofer pattern\cite{Bastian} in a planar
Josephson junction in external magnetic field. Note that in the
absence of Zeeman field ($h=0$), all spin effects disappear:
$\gamma(0,\bbox{w}) = 0$.

The most striking manifestation of the SO interaction itself is
the anisotropy of Andreev levels with respect to the direction of
the Zeeman field $\bbox{h}$; in the absence of the SO interaction
($\bbox{w} = 0$), $E_n(\Phi)$ depends only on the modulus of
$\bbox{h}$:
\begin{equation}
\gamma(\bbox{h},0) = \arcsin\left[\sin\left( hd/\hbar
v\right)\right].
\end{equation}
It is helpful to consider this anisotropy for a single electron
state with fixed direction of the SO vector $\bbox{w}$. As follows
from Eqs.\ (14)-(16), the energy of Andreev level depends only on
the angle between $\bbox{h}$ and $\bbox{w}$ and their moduli,
being insensitive to the rotation of $\bbox{h}$ around $\bbox{w}$.
At $\bbox{h} \parallel \bbox{w}$, the effects of SO interaction
vanish, as in Eq.\ (17). These conclusions can be extended to the
angle dependence of the Josephson current in Eq.\ (1) in a narrow
2DEG channel which holds a single electron mode ($p = p_F$, $p_y =
0$). In this extreme case, the vectors $\bbox{w}$ of all electrons
are directed along the $y$-axis and create a fixed reference frame
for the Zeeman vector $\bbox{h}$.

At arbitrary length of the 2DEG region, Eq.\ (14) can be solved
only numerically. Below we consider an analytically solvable case
of 2DEG channel much shorter than the coherence length $\xi_0$,
when the left-hand side of Eq.\ (14) is negligibly small,
\begin{equation}
E_\lambda(\Phi) = s\Delta\cos(\Phi/2 + \sigma\gamma), \qquad
(s,\sigma)=\pm 1.
\end{equation}

At $\gamma = 0$, Eq.\ (18) describes two spin-degenerated Andreev
levels in a superconducting constriction\cite{Furusaki} which
traverse across the whole energy gap with the change of $\Phi$ and
intersect each other at $\Phi = \pi$.\cite{foot2} The spin effects
split each level into two spin-dependent terms and, in addition,
expand them into four energy bands which transfer the Josephson
current
\begin{eqnarray}
\nonumber &\displaystyle I(\Phi) = {e\Delta\over\hbar}\!
\int^{p_F}_{-p_F}\!{dp_y\over 2\pi\hbar}\! \sum_{\sigma=\pm 1}
\tanh\!\left[{\Delta\over 2T}\cos\!\left({\Phi\over 2}+
\sigma\gamma \right)\!\right]
\\  &\displaystyle \times\sin\left(\Phi/ 2+\sigma\gamma\right).
\end{eqnarray}
At $h=0$, Eq.\ (19) is reduced to a 2D analogue of the
current-phase relationship\cite{KO} for pure constriction.

The set of curves $I(\Phi)$ calculated numerically at $T=0$ for
various directions and magnitudes of Zeeman field combined with SO
interaction is presented in Fig.\ 2b-d in dimensionless variables
\begin{equation}
\bbox{H} = (\hbar v_F/d)\bbox{h}, \qquad  \bbox{W} = (\hbar
v_F/d)\bbox{w},
\end{equation}
in comparison with those plotted for $W = 0$ in Fig.\ 2a. The
common features of these dependencies are drastic variations of
the shape of $I(\Phi)$ at $H\sim 1$, and the rapid change of sign
of the derivative $dI/d\Phi$ at $\Phi=\pi$ in small field $H$. It
is interesting that the curves at $W = 0$ and at $W = 1$,
$\bbox{H} \parallel \bbox{y}$ (Fig.\ 2a,b) are similar each to
other, as well as the curves for $W = 1$, $\bbox{H} \parallel
\bbox{x}$ and $W = 1$, $\bbox{H} \parallel \bbox{z}$ (Fig.\ 2c,d).
This reflects the results of our analysis of the anisotropy of
Andreev levels in 1D case which appears to be qualitatively
applicable for 2D system: the SO effects are relatively small at
$\bbox{H} \parallel \bbox{y}$ and approximately isotropic under
rotation of Zeeman field around the $y$-axis.

\begin{figure}
\epsfxsize=8.5cm\epsffile{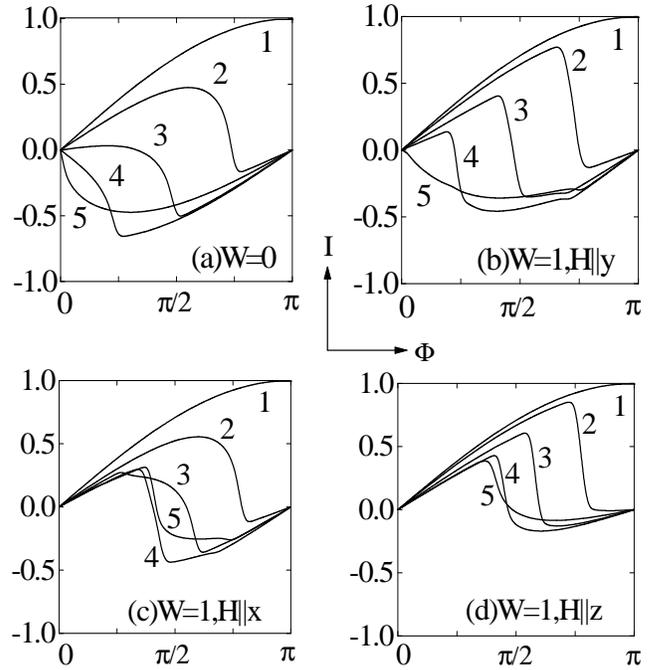} \vspace{3mm}
\caption{Current-phase relationships $I(\Phi)$ at $T=0$ normalized
on the critical current at $H=0$ for various directions and
magnitudes of the Zeeman field and intensities of the SO
interaction: $H=0$ (1), 0.4 (2), 0.8 (3), 1.2 (4), 1.6 (5). }
\end{figure}

\begin{figure}
\epsfxsize=8.5cm\epsffile{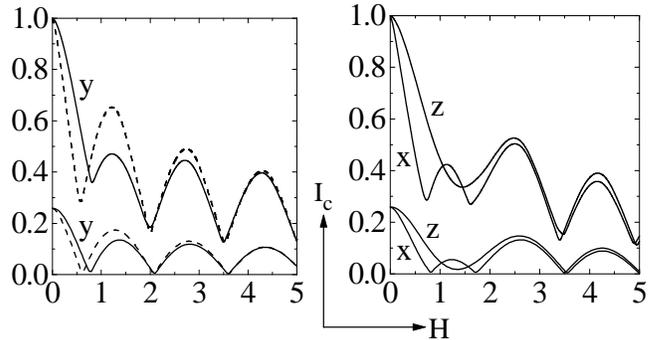} \vspace{3mm} \caption{ The
set of dependencies of the critical current
$I_c(\bbox{H},T)/I_c(0,0)$ on the dimensionless magnetic field
$H$, at $W=1$, $\bbox{H} \parallel \bbox{x}$, $\bbox{y}$,
$\bbox{z}$ (solid curves) and $W=0$, $\bbox{H}$ directed
arbitrarily (dashed curves). Upper pairs of curves were calculated
for $T=0$, lower pairs -- for $T=0.9T_c$.}
\end{figure}

In Fig.\ 3 we present oscillations of the critical current
$I_c(H)$ depending on the HF field direction and SO interaction
intensity, for two values of temperature. At the vicinity of
$T_c$, where $I(\Phi)$ has a single harmonic,
\begin{equation}
I(\Phi)\approx I_m\sin\Phi, \;\; I_m = {\pi e\Delta^2\over
(2\pi\hbar)^2T_c} \int_{-p_F}^{p_F} dp_y \cos 2\gamma,
\end{equation}
the critical current $I_c(W,H)=|I_m|$ turns to zero periodically
with $H$ like in SFS system.\cite{BB}

The positive sign of $dI/d\Phi$ at $\Phi=\pi$, which occurs at
$H\neq 0$ (Fig.\ 2), means that this state can be stable and may
produce persistent current in the ground state of a
superconducting loop with high enough inductance
($\pi$-junction\cite{pi}). On the other hand, the negative sign of
$dI/d\Phi$ at $\Phi = 0$, which occurs within the certain field
range at $W = 0$ or at $W \neq 0$, $\bbox{H} \parallel \bbox{y}$,
signifies instability of usual ground state with $\Phi = 0$ (note
that the SO interaction stabilizes this state at $\bbox{H} \perp
\bbox{y}$, at least at $T=0$). The results of a numerical analysis
of stability of states $\Phi=0$, $\pi$ within the whole
temperature range are shown in Fig.\ 4.

\begin{figure}
\epsfxsize=8.5cm\epsffile{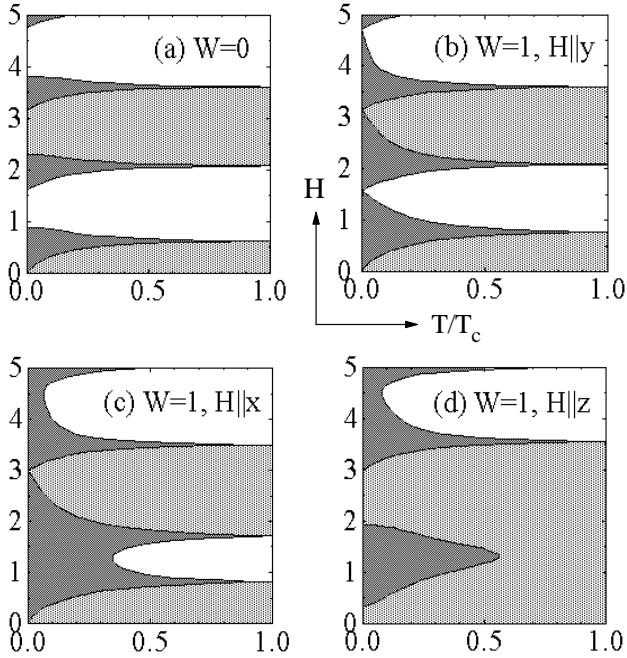}\vspace{2mm} \caption{
$H-T/T_c$ diagrams of stability of S-2DEG-S phase states with
$\Phi = 0$ and $\Phi=\pi$  for various directions and magnitudes
of the Zeeman field and intensities of the SO interaction. Within
the gray regions, only usual 0-state is stable: $dI/d\Phi(0) > 0$,
$dI/d\Phi(\pi) < 0$; blank regions correspond to the stability of
the $\pi$-state only: $dI/d\Phi(0) < 0$, $dI/d\Phi(\pi) > 0$;
within the dark regions both derivatives are positive. }
\end{figure}

In summary, we have shown that the Josephson current in a
mesoscopic S-2DEG-S structure is highly sensitive to the combined
action of the Zeeman field and spin-orbit interactions. In
particular, the critical current reveals oscillations and
anisotropy with respect to the Zeeman field $\bbox{B}$, and the
regions of stability at $\Phi =\pi$ (like in $\pi$-junctions)
emerge. We assumed hyperfine interaction of electrons with
polarized nuclei as the source of electron spin polarization,
though similar effects should be observed in external magnetic
field lying in the 2DEG-plane (to avoid orbital effects). In order
to access the regime of strong interaction ($H\sim 1$, $W \sim 1$)
in short 2DEG bridge ($d < \xi_0$) considered here, the
interaction energies $h$, $w$ of 2DEG should exceed $\Delta$.
Since the HF and SO interaction magnitudes in GaAs/AlGaAs
heterojunctions reach at most 1K in temperature scale, the banks
of short S-2DEG-S structure should be preferably fabricated from a
superconductor with low $T_c\leq 1$K. This restriction can be
significantly softened in long ($d \gg \xi_0$) S-2DEG-S junctions
where the interaction energies should be comparable with the small
distance between Andreev levels: $h,w \sim \hbar v_F/d \ll
\Delta$.


\begin{references}
\bibitem{foot1} Since the energy of Andreev ``levels'' in Eq.\ (14) depends on
the continuous electron quantum number $v(p_y)$, they actually
represent overlapping energy bands.
\end{references}
\end{document}